\documentstyle[aps,prl,epsf]{revtex}

\begin{document}
\twocolumn[\hsize\textwidth\columnwidth\hsize
\csname @twocolumnfalse\endcsname

\title{Coherent versus Incoherent $c$-Axis Josephson Tunneling
Between Layered Superconductors}
\author{R. A. Klemm,$^1$ G. Arnold,$^2$ C. T. Rieck and K. Scharnberg$^3$}
\address{$^1$Materials Science Division, Argonne National Laboratory,
Argonne, IL
60439 USA}
\address{$^2$Department of Physics, University of Notre Dame, Notre
 Dame, IN 46556}
\address{$^3$Fachbereich Physik, Universit{\"a}t
Hamburg, Jungiusstra\ss e 11, D-20355 Hamburg, Germany}

\date{\today}
\maketitle

\begin{abstract}
We calculate $I_c(T)$ and $R_n$ for both coherent and incoherent
 electron
tunneling
across a $c$-axis break junction between two $\nu = s,
d_{x^2-y^2}$-wave
layered superconducting half spaces, each with $c$-axis bandwidth
 $2J$.
Coherent quasiparticle tunneling only occurs for voltages $V<2J/e$,
leading to difficulties in measuring $R_n$ for underdoped samples.
 The coherent  part of $I_c(0)$ is
independent of
$\Delta_{\nu}(0)$ for $J/\Delta_{\nu}(0)<<1$, and can be large.
	Our results are discussed with regard to recent
experiments.
\end{abstract}
\vskip0pt
\vskip0pt\vskip0pt
]
\narrowtext

 It is
presently possible to prepare high quality $c$-axis Josephson
junctions of Bi$_2$Sr$_2$CaCu$_2$O$_{8+\delta}$ (BSCCO).
\cite{Bozovic,Zasadzinski,Li} Early thin film atomic layer-by-layer
 molecular beam epitaxy (ALL-MBE) preparations of trilayer
junctions of BSCCO separated by a thin layer of Dy-doped BSCCO
 were found to have  values of the product $I_cR_n$ of the critical
current times the normal resistance at temperature $T=0$
consistently about 0.5 mV, with $I_c$ and $R_n$  varying greatly.
\cite{Bozovic}	Recently, they	prepared a   single Josephson
junction within a single unit cell of  underdoped BSCCO, and
reported  $I_cR_n\approx5-10$ mV.\cite{Bozovic}  Also, a
blunt point contact
 tip  pressed onto a BSCCO
 surface  often resulted in
 an apparent $c$-axis break junction. \cite{Zasadzinski}    Overdoped
junctions typically had
  $I_cR_n\approx 2.4$ mV, well below the
Ambegaokar-Baratoff (AB) result.\cite{AB}  However, two
underdoped break junction samples had
  $I_cR_n(0)\approx
 15 -25$ mV, apparently in {\it excess} of the expected
BCS-derived AB result.
  Furthermore, exceedingly clean
$c$-axis break junctions
 were prepared by cleavage and subsequent refusion of BSCCO,
with or without a
 twist about the $c$-axis.\cite {Li}	Although $I_c(T)R_n$
was not presented in
 these works, we anticipate that such measurements will be
 available shortly.

Here we consider  tunneling across an untwisted $c$-axis
 break (or single intrinsic Josephson)
junction which is much	less conductive  than the bulk, intrinsic
 junctions
 between neighboring
CuO$_2$ layer pairs.  We find that for underdoped samples,
standard
measurements of $R_n$ at voltages $V>2\Delta(0)/e$, where
$\Delta(T)$ is
the superconducting order parameter (OP) amplitude, can be
unreliable, since they do not
 measure the coherent processes, which can dominate at $V=0$.
 Hence, the
apparent large values of $I_cR_n$ reported for underdoped
samples could be
questionable.\cite{Bozovic,Zasadzinski}

 We assume a $c$-axis break junction between two untwisted
half-spaces of cross-sectional area $A$, each consisting of
$N>>1$ identical
clean superconducting layers separated a distance $s$ apart.
We label the
upper ($u$) and lower ($\ell$) half-spaces by $\mu=u,\ell$,
and  index the
layers in each half-space by $j=1,\ldots,N$, with $j=1$ being
the layer in each
half-space adjacent to the break junction.
 We allow $\psi_{\mu,j,\sigma}({\bf k}) \>
[\psi^{\dag}_{\mu,j,\sigma}
 ({\bf k})]$ to annihilate [create] a quasiparticle
 with spin $\sigma=\pm 1$
 and two-dimensional (2D) wavevector ${\bf k}$
 on the $j^{\rm th}$ layer within the ${\mu}^{\rm th}$
layered half-space.
 Within each layer in the $\mu^{\rm th}$ half-space, the
quasiparticles propagate with energy dispersions
$\xi_{\mu 0}({\bf k})=\epsilon_{\mu 0}({\bf k})-E_F$ relative
to the Fermi
energy $E_F$ [for
 free particles, $\epsilon_{\mu 0}({\bf k})=
 {\bf k}^2/(2m_{\mu})$], and interact with intralayer BCS-like
 pairing
interaction $ \lambda_{\mu}({\bf k},{\bf k}')=\lambda_{\mu\nu
0}\varphi_{\nu}(\phi_{\bf k})\varphi_{\nu}(\phi_{{\bf k}'})
$,  where $\nu=s,d$, $\varphi_s(\phi_{\bf k})=1$ and
$\varphi_d(\phi_{\bf
k})=\sqrt2\cos(2\phi_{\bf k})$ are the eigenfunctions
for $s$ and
$d_{x^2-y^2}$-wave intralayer  pairing, respectively.
We only
consider
here  the purely $s$-wave and $d$-wave cases
$\lambda_{\mu s 0}\ne\lambda_{\mu' d 0}=0$ and
$\lambda_{\mu d 0}\ne
\lambda_{\mu' s 0}=0$.  Between neighboring layers
in the $\mu^{\rm th}$ half-space, the quasiparticles
tunnel with matrix element $J_{\mu}/2$.\cite{KLB}
  The $c$-axis resistivity $\rho_c(T)$ above the transition
temperature $T_c$ suggests the limits $J_{\mu}/T_c>>1$
and $J_{\mu}/T_c<<1$ apply to overdoped and underdoped
materials, respectively.\cite{KM}

 In addition, we take the single particle tunneling  Hamiltonian
 $H_{\cal T}$ across the break junction to be
 \begin{equation}
 H_{\cal T}={1\over{A^2}}\!\!\sum_{{\bf k},{\bf k}',\sigma}\!\!
{\cal T}_{{\bf k},{\bf k}'}\psi^{\dag}_{u,1,\sigma}({\bf
 k})\psi_{\ell,1,\sigma}({\bf k}')+H. c.,\label{HT}
 \end{equation}
which transfers a quasiparticle from the
 $j=1$ layer in the $\ell$ half-space to the $j=1$ layer in
 the $u$ half-space, and {\it vice versa}; ${\cal T}_{{\bf k},
{\bf k}'}={\cal
T}^{*}_{{\bf
 k}',{\bf k}}$.  We set $\hbar=c=k_B=1$.

 We assume  both coherent and
incoherent
break junction tunneling.  The {\it spatially constant coherent}
 tunneling preserves the intralayer
 wavevectors, ${\bf k}={\bf k}'$.  However, pure {\it spatially
 random incoherent}  tunneling assumes ${\bf
 k}$ and ${\bf k}'$ are {\it independent} of each other, \cite{AB}
 which
allows no
$d$-wave  incoherent Josephson tunneling.  Hence, to allow
for a finite
(albeit
extremely small) amount of $d$-wave incoherent Josephson
tunneling, we
assume to second order in ${\cal T}_{{\bf k},{\bf k}'}$,
\cite{KRS,Sauls}
 \begin{eqnarray}
 \Bigl<{\cal T}_{{\bf k},{\bf k}'}{\cal T}_{{\bf k}',{\bf
 k}''}\Bigr>&=&A\delta_{{\bf k},{\bf
 k}''}\Bigl[|{\cal T}_0|^2A\delta_{{\bf k},{\bf k}'}
 +f_{\rm inc}({\bf k}-{\bf
 k}')\Bigr],\label{dwavetunneling}
 \end{eqnarray}
  where
 \begin{equation}
 f_{\rm inc}({\bf k}-{\bf
 k}')={{1}\over{2\pi N_{2D}(0)}}\!\left[ {1\over{\tau_{\perp s}}}\!+\!
 {{2\cos[2(\phi_{\bf k}\!-\!\phi_{{\bf
 k}'})]}\over{\tau_{\perp d}}} \right]\label{finc}
 \end{equation}
$1/\tau_{\perp d}<<1/\tau_{\perp s}$, and  $N_{2D}(0)=
[N_{2Du}(0)N_{2D\ell}(0)]^{1/2}$ is the geometric mean
 2D density of states; for
 free particles, $N_{2D}(0)=\overline{m}/(2\pi)$, where
$\overline{m}=(m_um_{\ell})^{1/2}$.
   In Eq.
 (\ref{dwavetunneling}),  $<\cdots>$ denotes a 2D
 spatial average.

 For intralayer pairing in a half-space with one conducting
 layer per unit
 cell, the regular and anomalous temperature Green's functions
$G_{\mu,j}({\bf k},\omega)$
 and $F_{\mu,j}({\bf k},\omega)$, where $\omega$  represents
 the Matsubara frequencies,  for propagation  within layer
 $j$ in the $\mu^{\rm th}$
half-space, explicitly depend upon $j$.\cite{LiuKlemm,KS}
However, the
OP
$\Delta_{\mu}$	is independent	of $j$.\cite{LiuKlemm}
 Nevertheless, $\Delta_{\mu}(\phi_{\bf
k})\equiv\Delta_{\mu\nu}(T)\varphi_{\nu}(\phi_{\bf k})$
implicitly
depends
upon  $\nu= s, d$.\cite{KL}

For purely $s$-wave incoherent tunneling between 3D
superconductors,
$I_cR_n$ is independent of the
properties of the junction \cite{AB}.
 However, in our model these
quantities must be evaluated separately.
$R_n$ is found from
the  quasiparticle  current $I_{qp}$ to leading order in
${\cal T}_{{\bf k},{\bf k}'}$,\cite{Duke}
\begin{eqnarray}
 I_{qp}&=&{{4e}\over{A^2\pi}}\sum_{{\bf k},{\bf k}'}
\Bigl<|{\cal T}_{{\bf k},{\bf
 k}'}|^2\Bigr>\int_{-\infty}^{\infty}d\epsilon[f(\epsilon_u)-
f(\epsilon_{\ell})]
\times\cr
& &\cr
& &\qquad\times\Im[G_{u,1}({\bf k},-i\epsilon_u)]
\Im[G_{\ell,1}
({\bf k}',-i\epsilon_{\ell})],
 \end{eqnarray}
where $f(x)$ is the Fermi function,
$\epsilon_u=\epsilon$, $\epsilon_{\ell}=\epsilon+eV$,	and
the $G_{\mu,1}({\bf k},-i\epsilon_{\mu})$ are obtained from
 $G_{\mu,1}({\bf k},\omega)$ by the
analytic continuations $\omega\rightarrow-i\epsilon_{\mu}$.
Since the tunneling takes place between the $j=1$ in
the two half-spaces, the only relevant wavevectors are 2D. Hence, we set
$\sum_{{\bf k}}\rightarrow AN_{2D}(0)\int_{-\infty}^{\infty}
d\xi_{\mu 0}\int_0^{2\pi}d\phi_{{\bf k}}/(2\pi)$.

We  consider separately the coherent and incoherent processes,	and
separately the $G_{\mu,1}({\bf k},\omega)$
as evaluated exactly for the layered half-spaces, and as
 {\it approximated} using
the bulk layered states. In the  bulk-space treatment, we assume
$G_{\mu,1}({\bf k},\omega)\approx G_{\mu,b}({\bf k},\omega) \allowbreak
=\int_0^{\pi}\frac{dz}{\pi}/[i\omega-\xi_{\mu 0} - J_{\mu}\cos z]$,
or $G_{\mu,b}({\bf k},\omega)=1/R_{\mu}(i\omega)$, where
$R_{\mu}(z)\equiv[(z-\xi_{\mu 0})^2-J_{\mu}^2]^{1/2}$ depends upon
${\bf k}$ only through $\xi_{\mu 0}({\bf k})$.
\cite{KLB}
However, when one properly takes account of the surface at the weak break
junction, \cite{LiuKlemm} $G_{\mu,1}=\allowbreak
\int_0^{\pi}\frac{2dz}{\pi}\sin^2z
/[i\omega-\xi_{\mu 0} - J_{\mu}\cos z]$, or
$G_{\mu,1}({\bf k},\omega)=\Xi_{\mu}(i\omega)$,
where
\begin{equation}
\Xi_{\mu}(z)=2/[z-\xi_{\mu 0}+R_{\mu}(z)].\label{Xi}
\end{equation}
Using either the bulk or half-space states and the identity
$\int_{-\infty}^{\infty}d\epsilon[f(\epsilon)-f(\epsilon+eV)]
 = eV$, the incoherent quasiparticle $c$-axis break junction
 tunneling current is Ohmic,
\begin{equation}
I^{\rm inc}_{qp}=V/R_n^{\rm inc}=2e^2VN_{2D}(0)
/\tau_{\perp s}.\label{Ispinc}
\end{equation}

For the coherent   quasiparticle $c$-axis break junction
tunneling current, we only consider tunneling between
identical materials with $J_{\ell}=J_{u}=J$, etc., and obtain
\begin{equation}
I_{qp}^{\rm coh}(V)=V/R_n^{\rm coh}(V)=64C/(3\pi)\gamma
\Theta(1-\gamma^2)Q(\gamma),
\label{Ispcoh}
\end{equation}
where  $C=2e|{\cal T}_0|^2N_{2D}(0)$,
$Q(\gamma)=(1+|\gamma|)[(1+\gamma^2)E(k)-2|\gamma|K(k)]$,
$k=\left(1-|\gamma|\right)/(1+|\gamma|)$, $\gamma=eV/(2J)$,
 $K(z), E(z)$
are
standard complete elliptic integrals, and $\Theta(z)$ is the
Heaviside step
function. In the bulk state approximation, $Q(\gamma)$ is
replaced by
$(3/8)K(k)/(1+|\gamma|)$, which leads
to a spurious,
non-Ohmic $\ln{V}$ dependence of $1/R_n^{\rm coh}$ as
$V\rightarrow0$.

Although $I_{qp}^{\rm coh}(V)$ is Ohmic for $V\rightarrow 0$,
it is non-Ohmic for finite $V$, and {\it vanishes}
for $|eV|\ge 2J$.  Thus, for
$|eV|\ge2J$, the
only quasiparticle
 tunneling across the break junction process allowed
is {\it incoherent}.  This result  arises from
the geometry:  the half-spaces are assumed to be {\it layered},
 each with $c$-axis bandwidth $2J$.  For 3D
systems, there is no such limitation upon $I_{qp}^{\rm coh}$.

Thus, the quasiparticle current consists of two parts,
 $I_{qp}(V)=I_{qp}^{\rm inc}+I_{qp}^{\rm coh}(V)=V/R_n(V)$.
 In the inset of Fig. 1, we have plotted $R_0/R_n(V)$, where
$R_0=J/[8\pi^2e^2N_{2D}(0)|{\cal T}_0|^2]$,
as a function of $eV/2J$.  Note that one requires the break junction
conductance to be much less that the conductance across neighboring
layers in each half space. This implies that both  $1/\tau_{\perp s}$ and
$|{\cal T}_0|^2/J$ are	small with respect to $J^2\tau_{||}$, where
$1/\tau_{||}<<J$ is the small intralayer scattering rate.
\cite{KRS}  However, this does not restrict  the relative
magnitudes of $I_{qp}^{\rm coh}$ and $I_{qp}^{\rm inc}$.

\begin{figure}[htb]
\vspace*{-1.2cm}
\epsfxsize=9cm
\centerline{\epsffile{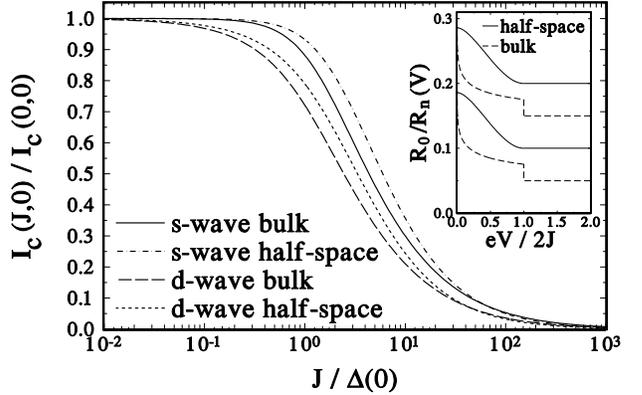}}
\caption{$I^{\rm coh}_{c,\nu}(J,0)/I^{\rm coh}_{c,\nu}(0,0)$
 and  $I^{\rm coh}_{c,\nu,b}(J,0)/I^{\rm coh}_{c,\nu,b}(J,0)$ are plotted
versus	$\log_{10}[J/\Delta_{\nu}(0)]$ for $\nu= s, d$.
Inset: Plot of $R_0/R_n(V)$, where $R_0=J/[8\pi^2e^2
N_{2D}(0)|{\cal T}_0|^2]$, and its bulk space approximation
versus $eV/2J$. The curves are shifted by  $J/[4\pi^2|{\cal
T}_0|^2\tau_{\perp s}]$ values of 0.05, 0.1, 0.15, and 0.2
for clarity.}
\end{figure}

In the superconducting state, $I_c$  for Josephson  tunneling across the
break junction	between arbitrary layered half spaces is given to lowest
order  in ${\cal T}_{{\bf k},{\bf k}'}$ by \cite{KRS}
 \begin{equation}
 I_c(T)=\frac{4eT}{A^2}\!\!\sum_{\omega,{\bf k},{\bf k}'}\!
 \left<\!|{\cal T}_{{\bf k},{\bf k}'}|^2\!\right>F_{u,1}({\bf
 k},\omega)F_{\ell,1}^{\dag}({\bf k}',\omega),
\label{Josephson}
 \end{equation}
where quite generally
$ F_{\mu,1}=-\Delta_{\mu}(\phi_{\bf k})\Im \Gamma_{\mu}/D_{\mu}$,
$\Gamma_{\mu}=\left[\exp\left(ik_{+}s\right)-\exp\left(ik_{-}s\right)\right]/
(iJ_{\mu})$ and  $D_{\mu}\equiv[|\Delta_{\mu}|^2+\omega^2]^{1/2}$.
\cite{LiuKlemm}
The quantities $\exp(ik_{\pm}s)$ are obtained from the equation
$J_{\mu}\cos(k_{\pm}s)=-\xi_{\mu 0}({\bf k})\pm iD_{\mu}$,
\cite{LiuKlemm}
which leads to	$\Gamma_{\mu}=\Xi_{\mu}(iD_{\mu})$,
where $\Xi_{\mu}(z)$ is given by Eq. (\ref{Xi}).
In the bulk state approximation for $F_{\mu,1}$, $\Gamma_{\mu}$ is replaced by
$R^{-1}_{\mu}(iD_{\mu})$. Note that as $J_{\mu}\rightarrow0$,
the half-space and bulk expressions both reduce to the familiar 2D form.

The incoherent part of the break junction $I_c$ between arbitrary layered
superconductors  with $\nu= s, d$ is
\begin{equation}
I^{\rm inc}_{c,\nu}(T)={{2eN_{2D}(0)T\pi}
\over{\tau_{\perp\nu}}}\sum_{\omega}
\prod_{\mu=u}^{\ell}\int_0^{2\pi}\!
{{d\phi_{\bf k}\Delta_{\mu}\varphi_{\nu}}\over{2\pi D_{\mu}}}.
\label{incoherent}
\end{equation}
Note that $\Delta_{\mu}(\phi_{\bf
k})$ implicitly depends upon $\nu$.  Equation
(\ref{incoherent}) is
obtained using either the bulk or half-space states.
For $s$-wave pairing,
$I_{c,s}^{\rm inc}(T)R^{\rm inc}_n$ equals  the AB result.
\cite{AB}
For $\Delta_{u}=\Delta_{\ell}=\Delta$,
$I_{c,\nu}^{\rm inc}(0)=\pi eN_{2D}(0)\Delta_{\nu}(0)
A_{\nu}/\tau_{\perp\nu}$, where $A_s=1$, and
 $A_d=0.153$.  For this case,
$I_{c,s}^{\rm inc}(T)/I_{c,s}^{\rm inc}(0)$ and  $I_{c,d}^{\rm
inc}(T)/I_{c,d}^{\rm inc}(0)$ are plotted in Figs. 2 and 3,
respectively.  We note that
$I^{\rm inc}_{c,d}(T)/I_{c,s}^{\rm inc}(T)
\propto\tau_{\perp s}/\tau_{\perp d}<<1$.

For coherent $c$-axis break junction Josephson tunneling between
 identical materials, we  drop the
subscripts $\mu$, noting that $\Delta(\phi_{\bf
k})=\Delta_{\nu}(T)\varphi_{\nu}(\phi_{\bf k})$, etc.
Writing $\Gamma$ in $F_1$ in integral form as above Eq. (\ref{Xi}),
 and performing two integrals
 analytically, we have
\begin{equation}
I^{\rm coh}_{c,\nu} (J,T)=\frac{4C T}{\pi} \sum_{\omega}
\int^{2 \pi}_0d\phi_{\bf k}|\Delta|^2\int^{\pi}_0
\frac{dzX(z)}{D^3
W(z)}\, ,\label{coherent}
\end{equation}
where $W(z)=\left[1+J^2\sin^2z/D^2\right]^{1/2}$, $X(z)=\sin^4z -
\sin^2z/[1 + W(z)]$,
  and $C=2e|{\cal
T}_0|^2N_{2D}(0)$. Using  the
bulk states, $I_{c,\nu,b}^{\rm coh}(J,T)$ is obtained from
 Eq. (\ref{coherent}) by
replacing $X(z)$ by $\sin^2z/8$.
 In the limit $T\rightarrow0$, we set $\alpha(\phi_{\bf k}) =
|J/\Delta_{\nu}(0)\varphi_{\nu}(\phi_{\bf k})|$, $\beta(\phi_{\bf k})={1\over2}(1+[1+\alpha^2]^{1/2})$, and
$I_{c,\nu}^{\rm coh}(J,0)=CY_{\nu}[J/\Delta_{\nu}(0)]$ reduces to
\begin{equation}
I^{\rm coh}_{c,\nu} (J,0) =\frac{4C}{\pi}\int^{2 \pi}_0
d\phi_{\bf k}
\left(\frac{1 + 8\beta}{24\beta^2}
- \frac{\ln\beta}{\alpha^2} \right). \label{equaldeltas}
\end{equation}
  For the bulk states, $I^{\rm
coh}_{c,\nu,b}(J,0)$ is obtained from
$I^{\rm coh}_{c,\nu}(J,0)$ by
replacing the quantity in large
brackets by $(8\alpha)^{-1}\sinh^{-1}\alpha$.  In Fig. 1,
we have plotted $I^{\rm
coh}_{c,\nu}(J,0)/I_{c,\nu}^{\rm coh}(0,0)$ and $I^{\rm
coh}_{c,\nu,b}(J,0)/I_{c,\nu,b}^{\rm coh}(0,0)$ versus
$\log_{10}[J/\Delta_{\nu}(0)]$ for
$s$- or $d$-wave OPs.
  The most surprising
point is
that for small $J/\Delta_{\nu}(0)$,  $Y_{\nu}(0)=1$, so that
$I_{c,\nu}^{\rm coh}(J,0)\rightarrow C$,
  {\it independent} of	$\Delta_{\nu}(0)$.

From Fig. 1, $I^{\rm coh}_{c,d}(J,0)$ is slightly more
 sensitive
to $J$
than is $I^{\rm coh}_{c,s}(J,0)$.  In addition, for intermediate
values of $J/\Delta_{\nu}(0)$, the
respective $s$- and $d$-wave bulk curves closely
approximate the half-space
curves obtained by reducing $J/\Delta_{\nu}(0)$ by
the constant factor $1/\sqrt2$.
For $J/\Delta_{\nu}(0)>>1$, however, there are distinct
 differences between the bulk
and half-space curves.	Whereas  the correct
$I_{c,\nu}^{\rm coh}=CY_{\nu}[J/\Delta_{\nu}(0)]
\rightarrow16C\Delta_{\nu}(0)B_{\nu}/(3J)$,
 where $B_s=1$, and $B_d=2\sqrt2/\pi$, the approximate
$I_{c,\nu,b}^{\rm coh}(J,0)$ spuriously reduces to
$CB_{\nu}[\Delta_{\nu}(0)/J]\ln[2J/D_{\nu}
\Delta_{\nu}(0)]$, where $D_s=1$, $D_d=\sqrt2/e$.

It is interesting to
compare the coherent and incoherent  results for identical
 half-spaces.  At $T=0$, the $V=0$ $I_{qp}^{\rm coh}(0)
/I_{qp}^{\rm inc}\propto|{\cal T}_0|^2\tau_{\perp s}/J$,
whereas for $T>T_c$, $J/T_c$ is the relevant quantity
for the half spaces.\cite{KM}
 Since $\Delta_{\nu}(0)\approx T_c$, for $J/T_c<<1$,
$I_{c,\nu}^{\rm coh}(J,0)/I_{c,\nu}^{\rm inc}(0)\propto
|{\cal
T}_0|^2\tau_{\perp\nu}/T_c$. For
$J/T_c>>1$,
   $I_{c,\nu}^{\rm coh}(J,0)$ and $I_{c,\nu}^{\rm inc}(0)$
 both $\propto\Delta_{\nu}(0)$, but
$I_{c,\nu}^{\rm coh}(J,0)/I_{c,\nu}^{\rm inc}(0)
\propto |{\cal
T}_0|^2\tau_{\perp\nu}/J$.  These results lead to the
curious conclusions that
for $J/T_c<<1$, the underdoped normal state tunneling
is incoherent, the $T=0$ quasiparticle break junction
tunneling could be either coherent or incoherent, but
$d$-wave break junction pair tunneling would be mainly
coherent.  On the other hand, for $J/T_c >>1$, the
overdoped normal state half-space tunneling would be
coherent, but the $T=0$
quasiparticle break junction tunneling and the $s$-wave
break junction pair tunneling could be incoherent!

  In the limits
$J_{\mu}\rightarrow0$,
one can evaluate the $J=0$, $T=0$ limit of the coherent
 part of $I_c(T)$
from Eq.
(\ref{Josephson}) as a function of
$r=\Delta_u/\Delta_{\ell}$,
obtaining
$I_c^{\rm coh}(0,0)=2Cr\ln(r)/(r^2-1)$,
where $C$ is given following Eq. (\ref{Ispcoh}).  In the
limit $r\rightarrow1$,
$I^{\rm coh}_c(0,0)\rightarrow C$.   For either two $s$-wave
 or two $d$-wave superconductors, $r$ is independent of
 $\phi_{\bf k}$.

In Fig. 2, we have plotted
$I_{c,s}^{\rm coh}(J,T)/I_{c,s}^{\rm coh}(J,0)$, as
 a function of
$T/T_c$, for tunneling between two layered $s$-wave
superconductors.  Typical
curves with $J/\Delta_s(0)=0,2,10$ are shown, along
with the AB curve, Eq.
(\ref{incoherent}).  For
$J/\Delta_s(0)=100$, the curve is almost identical to
 the AB curve.	Using the
bulk states doesn't change these curves very much, except
 for  large
$J/\Delta_s(0)$.	Clearly,  $I_{c,s}(J,T)/I_{c,s}(J,0)$ is
rather indistinguishable from that of AB, independent of
the microscopic
details.

\begin{figure}[htb]
\vspace*{-1.2cm}
\epsfxsize=9cm
\centerline{\epsffile{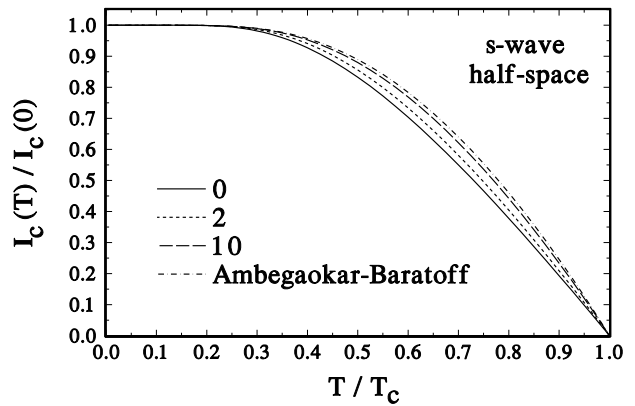}}
\caption{Plots of $I^{\rm coh}_{c,s}(J,T)
/I^{\rm coh}_{c,s}(J,0)$  for
$J/\Delta_s(0)=0, 2,10$ and of
 $I^{\rm inc}_{c,s}(T)/I^{\rm inc}_{c,s}(0)$ (AB)
versus $T/T_c$, for tunneling between identical
$s$-wave half-space  superconductors.}
 \label{fig2}
\end{figure}

In Fig. 3, we have plotted $I_{c,d}^{\rm coh}(J,T)
/I_{c,d}^{\rm coh}(J,0)$, as
a function of $T/T_c$.	Typical curves with
$J/\Delta(0)=0,0.5,2,10$ are shown along with  the
$d$-wave analog of AB, $I_{c,d}^{\rm
inc}(T)/I_{c,d}^{\rm inc}(0)$.	Note that  the	magnitude
 of $I^{\rm inc}_{c,d}(T)$
is  very small, due to the factor of $1/\tau_{\perp d}$.
Unlike the
$s$-wave curves in Fig. 2, the $I_{c,d}^{\rm coh}(J,T)
/I_{c,d}^{\rm coh}(J,0)$
curves with
small values of $J/\Delta_d(0)$ are distinctly linear at
low $T$, and are thus
distinguishable from the AB curve. in Fig. 2.  However,
for $J/\Delta_d(0)>>1$,   $I^{\rm coh}_{c,d}(J,T)
/I^{\rm coh}_{c,d}(J,0)$ and $I_{c,d}^{\rm inc}(T)
/I^{\rm inc}_{c,d}(0)$	are are nearly
indistinguishable from the AB curve.

\begin{figure}[htb]
\vspace*{-1.2cm}
\epsfxsize=9cm
\centerline{\epsffile{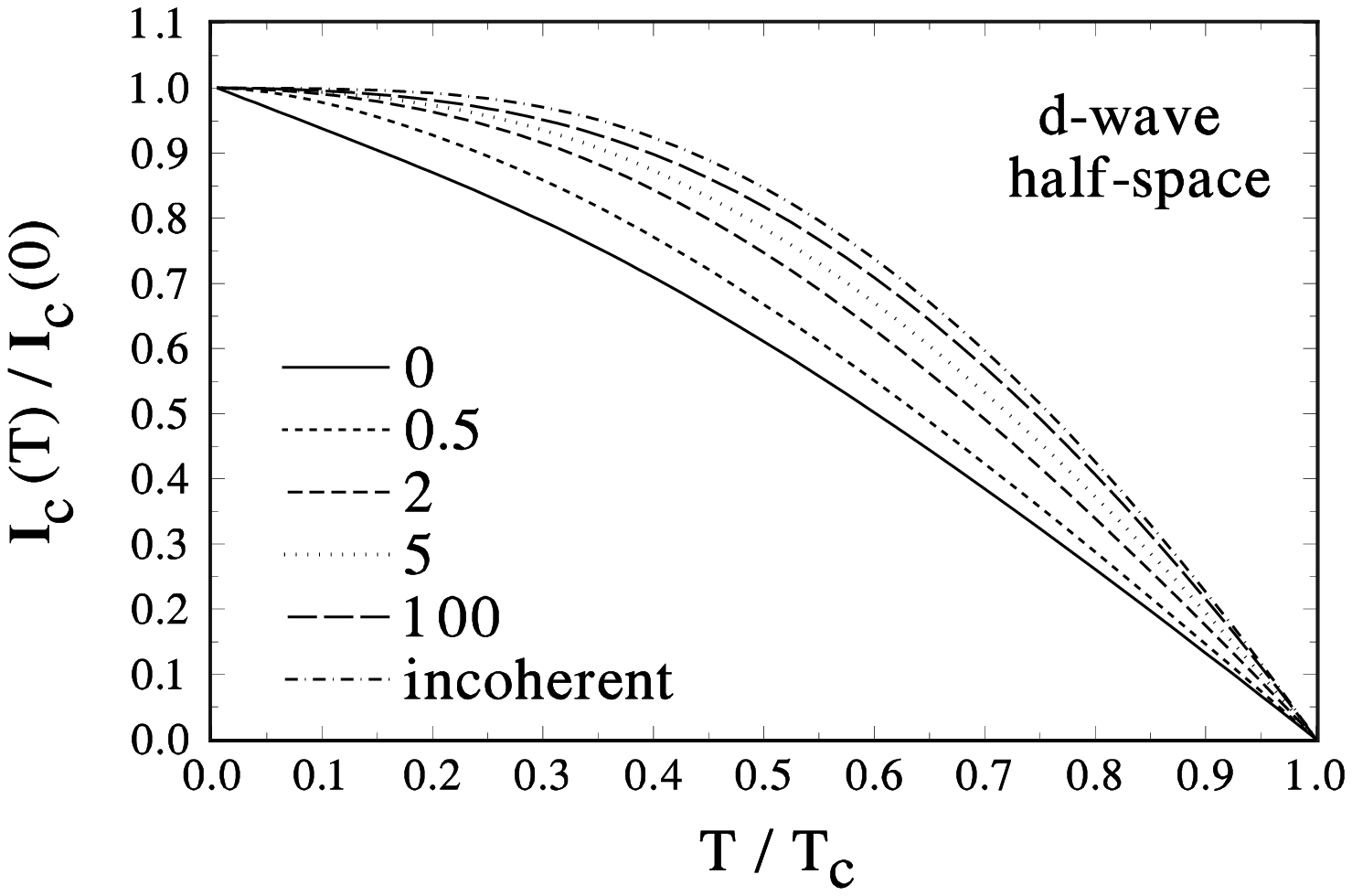}}
\caption{Plots of $I^{\rm coh}_{c,d}(J,T)
/I^{\rm coh}_{c,d}(J,0)$   for $J/\Delta_d(0)=0,
\allowbreak 0.5,\allowbreak 2,\allowbreak 5,
\allowbreak 100$ and of $I^{\rm inc}_{c,d}(T)
/I^{\rm inc}_{c,d}(0)$ versus $T/T_c$, for
tunneling between identical $d$-wave half-space
 superconductors.}
\end{figure}

Summarizing our results, we have for $\nu=s,d$,
\begin{equation}
I_{c,\nu}(J,0)R_n(0)=\frac{\pi[Z_{\nu}\Delta_{\nu}(0)
+3J\eta Y_{\nu}/16]}{2e[1+\eta]},\label{answer}
\end{equation}
where  $Z_{\nu}=A_{\nu}\tau_{\perp s}/\tau_{\perp\nu}$
 and $\eta=32|{\cal T}_0|^2\tau_{\perp s}/(3\pi J)$. Since
 $I_c$ is measured at $V=0$, one requires $R_n(0)$.  For
 overdoped  samples,
$J/T_c>>1$, and
$I_{c,\nu}(J,0)R_n(0)$ reduces to $\pi\Delta_{\nu}(0)
[Z_{\nu}+\eta B_{\nu}]/\{2e[1+\eta]\}$.  For both $\nu=s,d$,
this
is proportional to $\Delta_{\nu}(0)$, and nearly
 independent of the break
junction properties for $s$-wave superconductors.
 For $d$-wave superconductors, one requires a
substantial $I_{c,d}^{\rm coh}$ in order to obtain a
non-negligible
 $I_cR_n$.  However, for underdoped samples,
$J/T_c<<1$, and the situation is far more complicated.
 First of all, one
 cannot measure $R_n(0)$ in the usual way, since the
 coherent contribution,
which can be large at $V=0$, vanishes for
 $eV/2\Delta_{\nu}(0)>1$.  Moreover,
 $I_{c,\nu}(J,0)$  is dominated by coherent tunneling
and independent of $\Delta_{\nu}(0)$ for
$|{\cal T}_0|^2>>\pi A_{\nu}\Delta_{\nu}(0)
/2\tau_{\perp\nu}$, which is especially likely
for $d$-wave superconductors.

In conclusion, we found that for $c$-axis break junction
tunneling between two
layered superconductors, a crossover from incoherent
quasiparticle to
coherent pair tunneling can occur.  This  greatly
complicates the
determination of $R_n(0)$, the coherent part of which
cannot be seen from
measurements with $eV/2\Delta_{\nu}(0)>1$, unless
$J>>\Delta_{\nu}(0)$, which corresponds to overdoped
samples.  For underdoped samples, $I_cR_n$ values
 tend to be overestimated.
 The approximate bulk electronic states lead to
correct incoherent, but incorrect coherent, tunneling
results.
Incoherent $d$-wave pair tunneling leads only to very
small $I_cR_n$ values.
For coherent pair tunneling, both $s$-wave and $d$-wave
pair tunneling are
large in magnitude for small $J/\Delta_{\nu}(0)$, and
cross over to the AB form for
large $J/\Delta_{\nu}(0)$.  The $T$-dependence of
coherent $d$-wave tunneling is
distinctly different from that for AB for small
$J/\Delta_{\nu}(0)$.	Thus, accurate
measurements of the $T$-dependence and magnitude of
$I_cR_n$ in such break
junctions could give important information regarding the
questions of the order parameter
symmetry and  of the coherence of the pair
tunneling.


The authors  thank K. Gray, Q. Li, and J. F. Zasadzinski
 for useful discussions.
This work was supported by the
USDOE-BES through Contract No. W-31-109-ENG-38,
by NATO through Collaborative Research Grant No. 960102,
and by the DFG through
the Graduiertenkolleg ``Physik nanostrukturierter Festk\"orper.''

\end{document}